\documentclass[aps,prd,nofootinbib,twocolumn,floatfix,superscriptaddress,eqsecnum]{revtex4-1}
\usepackage{amsmath,amssymb,amsthm,bm,bbm}
\usepackage{graphicx,psfrag,subfigure,rotating}

\begin{document}
\title{Critical analysis of topological charge determination in the background
of center vortices in SU(2) lattice gauge theory}

\author{R. H\"ollwieser}
\author{M. Faber}
\affiliation{Atomic Institute, Vienna University of Technology, Wiedner Hauptstr.\ 8-10, A-1040 Vienna, Austria}

\author{U.M. Heller}
\affiliation{American Physical Society, One Research Road,
Ridge, NY 11961, USA}

\date{\today}
\begin{abstract}
We analyze topological charge contributions from classical $SU(2)$ center
vortices with shapes of planes and spheres using different topological charge
definitions, namely the center vortex picture of topological charge, a discrete
version of $F\tilde F$ in the plaquette or hypercube definitions and the lattice index theorem. For the latter the zero modes of the Dirac operator in the fundamental and adjoint representations using both the overlap and asqtad
staggered fermion formulations are investigated. We find several problems for
the individual definitions and discuss the discrepancies between the different
topological charge definitions. Our results show that the interpretation of
topological charge in the background of center vortices is rather subtle. 
\end{abstract}
\maketitle

\section{Introduction}
Since Savvidy~\cite{Savvidy:1977as} we know that the QCD-vacuum is non-trivial
and has magnetic properties. In lattice QCD it was shown~\cite{Greensite:2003bk}
that vortices (quantized magnetic fluxes) condense in the vacuum and compress the electric
flux between quark and antiquark to a string leading to confinement. This center
vortex
model~\cite{'tHooft:1977hy,Vinciarelli:1978kp,Yoneya:1978dt,Cornwall:1979hz2,Mack:1978rq,Nielsen:1979xu}
seems to be a very promising candidate to explain the phenomena that dominate
the infrared regime. Numerical simulations have indicated that vortices could
also account for phenomena related to chiral symmetry, such as causing
topological charge fluctuations and spontaneous chiral symmetry breaking
(SCSB)~\cite{deForcrand:1999ms,Alexandrou:1999vx,Engelhardt:2002qs,Hollwieser:2008tq}.
In particular,~\cite{Reinhardt:2000ck} states that the topological charge of a
vortex gauge field can be determined from the shape and orientation of P-vortices, {\it i.e.} from vortex intersections and writhing points.

Center Vortices are based on a discrete gauge symmetry of the action. A non-trivial center transformation of all link variables in one time (or space) slice~\cite{Greensite:2003bk}
\begin{equation}
U_0(\vec x,t_0)\Rightarrow zU_0(\vec x,t_0),\quad z\in Z_N
\end{equation}
leaves the action invariant. More generally, this transformation can be expressed
in terms of gauge transformations on a periodic lattice
\begin{equation}
U_0(\vec x,t_0)\Rightarrow g(x,t)U_0(\vec x,t_0)g^\dagger(x,t+1)
\end{equation}
which are periodic in the time direction only up to a $Z_N$ transformation:
\begin{equation}
g(\vec x,t_0+L_t)=zg(\vec x,t_0),\quad z\in Z_N
\end{equation}
This "singular" gauge transformation is not really a gauge transformation, since it affects of course the Polyakov loop, a gauge-invariant observable. Restricting such a transformation to a finite volume of a slice, the three dimensional Dirac volume, increases the action by a surface contribution, the vortex action. However, center vortices also increase the entropy, compensating the rise of the action and give essential
contributions to the vacuum configurations~\cite{Gubarev:2002ek}. The appearance
of link variables close to non-trivial center elements survives the continuum
limit and leads to singular gauge fields. This implies the question, whether
center vortices are lattice artifacts. One could give a positive answer, if
removing such “lattice artifacts” would not influence QCD. But it is just the
opposite, removing center vortices destroys confinement and the topological charge vanishes~\cite{deForcrand:1999ms}. This means center vortices are an essential ingredient of the QCD vacuum.

\section{Topology on the lattice}\label{sec:toplatt}

In lattice calculations there is a common method to determine the topological charge $Q_U$ from the integral
\begin{gather}
  Q_U = - \frac{1}{16\pi^2} \int d^4x \, Tr[\tilde{F}_{\mu\nu}
  F_{\mu\nu} ] \label{eq:qlatq},
\end{gather}
where $F_{\mu\nu}$ is expressed in terms of the plaquette field
\begin{gather}
P_{\mu\nu}=U_\mu(x)U_\nu(x+\mu)U_\mu^\dagger(x+\nu)U_\nu^\dagger(x).
\end{gather}
This expression is derived in the continuum from the transition between vacua with different winding numbers~\cite{Belavin:1975fg,Peskin:1978pa}
\begin{align}
  Q &= \int_{S_3} J_\mu d\sigma_\mu,\\
    &J_\mu=-\frac{1}{8\pi^2}\epsilon_{\mu\nu\alpha\beta} Tr(A_\nu\partial_\alpha
  A_\beta+2/3A_\nu A_\alpha A_\beta)\label{eq:qcont}.
\end{align}
Since $F_{\mu\nu}\tilde F_{\mu\nu}=\partial_\mu J_\mu$, $Q$ can be re-expressed
as the above volume integral~(\ref{eq:qlatq}). On the lattice, continuity in
space is lost and it seems that one should be able to view any lattice field
configuration as being a discrete copy of a smooth continuum configuration. This would always be topologically trivial, since $F\tilde F$ is a total
derivative. Nonzero $Q_U$ may come from field configurations containing gauge
singularities~\cite{Luscher:1981zq,Woit:1983tq,Fox:1984xd}.

Another possibility to analyze the topology of a gauge field is given by the
Atiyah-Singer index theorem
\cite{Atiyah:1971rm,Schwarz:1977az,Brown:1977bj,Narayanan:1994gw}. It states
that the topological charge of a gauge field configuration is proportional to
the index of the Dirac operator in this gauge field background. For the overlap Dirac
operator~\cite{Narayanan:1994gw,Neuberger:1997fp,Neuberger:1998wv} in the
fundamental representation the index is given by $\mathrm{ind}\;D[A] = n_- - n_+
= Q_D$, where  $n_-$ and $n_+$ are the number of left- and right-handed zero modes.
The adjoint version of the index theorem reads $\mathrm{ind}\;D[A] = n_- - n_+ =
2NQ_D = 4Q_D$, where $N=2$ is the number of colors and the additional factor $2$ is
due to the fact that the fermion is in a real representation, hence the
spectrum of the adjoint Dirac operator $iD$ is doubly degenerate. 
As described in~\cite{Hollwieser:2010mj} the improved staggered operator also
produces eigenmodes which can clearly be identified as zero modes and all results
in this paper show perfect agreement between the two fermion realizations,
considering that the eigenvalues of the staggered fermion operator have a
twofold degeneracy due to a global charge conjugation symmetry in $SU(2)$. We
therefore have $\mathrm{ind}\;D[A] = n_- - n_+ = 2Q_D$ for fundamental and
$\mathrm{ind}\;D[A] = n_- - n_+ = 8Q_D$ for adjoint (asqtad) staggered fermions.
The lattice version of the index theorem is only valid as long as the gauge
field is smooth enough and satisfies a so-called ``admissibility'' condition. It
requires that the plaquette values $U_{\mu\nu}$ are bounded close to $\mathbbm
1$, the value for very smooth gauge fields. Sufficient, but not necessary
bound for the ``admissibility'' of the gauge field are $|| 1 - U_{\mu\nu} || < 1
/ 30 $~\cite{Luscher:1981zq}, or $ || 1 - U_{\mu\nu} || < [6 (2+\sqrt(2))] = 0.04882 $~\cite{Neuberger:1999pz}. 

In this paper we discuss  smooth lattice field configurations, which fulfill the admissibility condition and show clearly a discrepancy between the integral of $F\tilde F$ and the topological charge derived from the lattice index theorem. These configurations are thick, spherical vortices in SU(2) lattice gauge theory. We analyze their topological charge and their zero modes. The problem seems to be related to the singular nature of vortex configurations, in fact, it is incorporated to the $SU(2)$ nature of our spherical vortex configuration. Here we analyze this problem in more detail and start with the discussion of plane vortices in different $U(1)$-subgroups of $SU(2)$.

\section{Plane Center Vortices and Topological Charge Contributions}\label{sec:plane}

\begin{figure}[htb]
\centering
\psfrag{z1}{\scriptsize $z_1$}
\psfrag{z2}{\scriptsize $z_2$}
\psfrag{2d}{\scriptsize $2d$}
\psfrag{1}{\scriptsize $1$}
\psfrag{0}{\scriptsize $0$}
\psfrag{p}{\scriptsize $\pi$}
\psfrag{2p}{\scriptsize $2\pi$}
\psfrag{12}{\scriptsize $N_z$}
\psfrag{f1}{$\phi$}
\psfrag{z}{$z$}
\includegraphics[width=.9\linewidth]{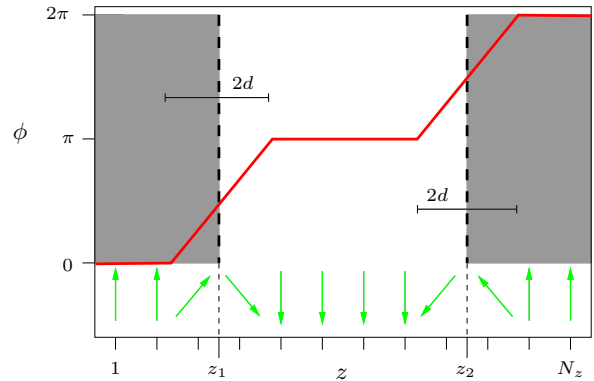}
\caption{The link angle $\phi$ of a parallel $xy$-vortex pair. The arrows ($t$-links) rotate clockwise with increasing $\phi$ in $z$-direction. The vertical dashed lines indicate the positions of the P-vortices. In the shaded areas the links have positive, otherwise negative trace.}
\label{fig:phis}
\end{figure}

For planar vortices parallel to two of the coordinate axes in
$SU(2)$ lattice gauge theory we analyzed the location of Dirac zero modes
in~\cite{Hollwieser:2011uj}. The vortices were defined by links varying in a
$U(1)$ subgroup of $SU(2)$, defined by the Pauli matrices $\sigma_i$, 
\begin{equation}
U_\mu=\exp(\mathrm{i} \phi \sigma_i).\label{eq:links}
\end{equation}
The direction of the flux and the orientation of the vortices were determined by the gradient of the angle $\phi$, which we choose as a piecewise linear function of the coordinate perpendicular to the vortex. The explicit functions for $\phi$ are given in Eq. (2.1) of~\cite{Hollwieser:2011uj}. For later use we plot in Fig.~\ref{fig:phis} the profile function $\phi$ and the corresponding $t$-links rotating in $z$-direction for a parallel $xy$-vortex.

Upon traversing a vortex sheet, the angle $\phi$ increases or decreases by $\pi$ within a finite thickness of the vortex. Center projection leads to a (thin) P-vortex at half the thickness~\cite{DelDebbio:1996mh}. If we consider these thick, planar vortices intersecting orthogonally, each intersection carries a topological charge with modulus $|Q|=1/2$, whose sign depends on the relative orientation of the vortex fluxes~\cite{Engelhardt:1999xw}. The plaquette definition simply discretizes the continuum (Minkowski) expression of the Pontryagin index to a lattice (Euclidean) version of the topological charge definition:
\begin{eqnarray}\label{equ:Qu}
  Q_U &=& - \frac{1}{16\pi^2} \int d^4x \, \mbox{tr}[\tilde{F}_{\mu\nu} {F}_{\mu\nu} ]\\
&=& - \frac{1}{32\pi^2} \int d^4x \, \epsilon_{\mu\nu\alpha\beta} \mbox{tr}[{F}_{\alpha\beta} {F}_{\mu\nu} ] = \frac{1}{4\pi^2} \int d^4x \, \vec E \cdot \vec B\nonumber
\end{eqnarray}
We build $xy$-vortices where only $zt$-plaquettes are non-trivial, i.e. with an electric field $E_z$, and $zt$-vortices bearing non-trivial $xy$-plaquettes corresponding to a magnetic field $B_z$. The topological charge is then proportional to $E_zB_z$. If the angle $\phi$ for different vortex sheets rotates in the same $U(1)$ subgroup of $SU(2)$, then parallel crossings give $Q=1/2$ and anti-parallel crossings give $Q=-1/2$. 

\begin{figure}
\psfrag{0}{}
\psfrag{x}{$x$}
\psfrag{z}{$z$}
\psfrag{+}{$+Q$}
\psfrag{Q}{}
\begin{tabular}{cc}
a)\includegraphics[keepaspectratio,width=0.45\linewidth]{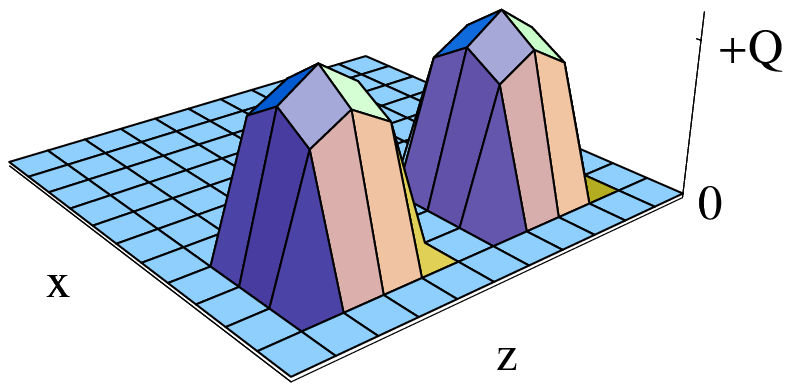} & \;b)\includegraphics[keepaspectratio,width=0.45\linewidth]{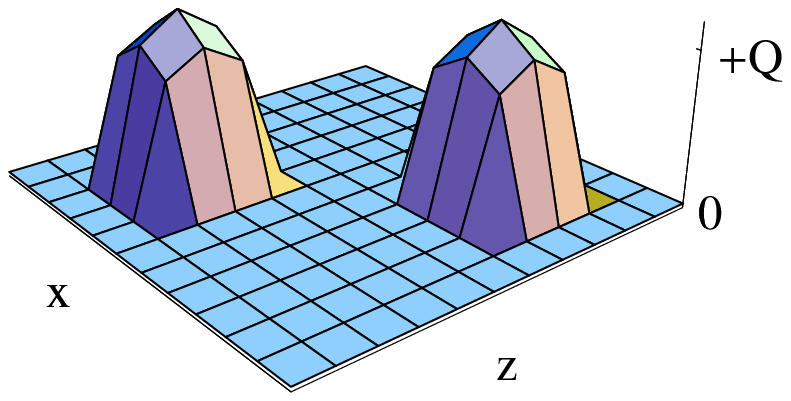}\\ c)\includegraphics[keepaspectratio,width=0.45\linewidth]{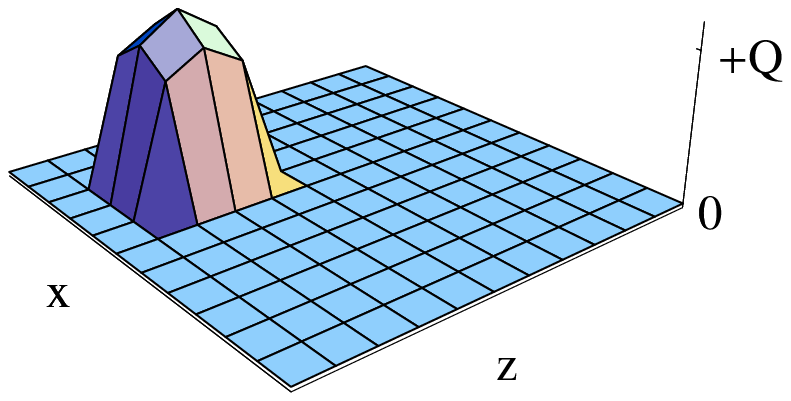} & \;d)\includegraphics[keepaspectratio,width=0.45\linewidth]{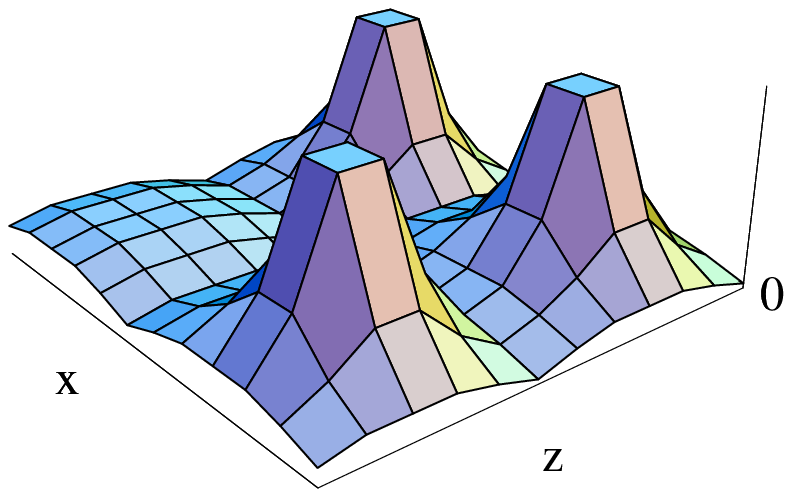} 
\end{tabular}
\caption{Topological charge density via $F\tilde F$ for two parallel vortices intersecting in four points (same geometry as in Fig. 1 of~\cite{Hollwieser:2011uj}  in the intersection plane. In all diagrams the first sheet of the $zt$-vortex rotates $\phi$ in $x$-direction from front to back from zero to $\pi$ in $\sigma_1$ and on to $2\pi$ in $\sigma_2$ for the second vortex sheet, whereas in (a) the $xy$-vortex rotates along $z$ from left to right in $\sigma_1$ only. In (b) the $xy$-vortex starts with $\sigma_2$ in the first sheet and continues with $\sigma_1$. In (a) and (b) we get $Q_U=1$. In (c) the $xy$-vortex starts with $\sigma_2$ in the first sheet and continues with $\sigma_3$ leading to $Q_U=1/2$. In (d) the action density in the intersection plane is shown for configuration (c).}\label{fig:qcontr} 
\end{figure}

In distinction to the above described discussion in
Ref.~\cite{Hollwieser:2011uj} we now change the $U(1)$ subgroup of $SU(2)$
in Eq.~(\ref{eq:links}) for the two crossing vortices. The profile function
$\phi$ remains the same as shown in Fig.~\ref{fig:phis}. The explicit
formula is again given by Eq.(2.1) in~\cite{Hollwieser:2011uj}. For an
orthogonal choice of $\sigma_i$ for $E_z$-plaquettes ($P_{zt}$) by non-trivial $U_t$ and $\sigma_j (i\ne j)$ for $B_z$-plaquettes ($P_{xy}$) by non-trivial $U_y$ the crossings do not contribute to the topological charge due to the orthogonality of the $U(1)$ subgroups. We note that,
because the $U_t$ and $U_y$ links are now noncommuting, also nontrivial
$P_{yt}$ plaquettes appear in the intersection region, but nontrivial
$P_{xz}$ plaquettes are absent, since the $U_x$ and $U_z$ links are trivial
everywhere, and so no contribution to $F\tilde F$ occurs. Maximal center
gauge still identifies an intersecting vortex pair. But the intersection
points do not carry topological charge $Q_U$ in the $F\tilde F$ definition
(\ref{equ:Qu}). For two intersecting vortex pairs we are able to construct
configurations with topological charge $Q_U=0$, $\pm 1$, and even $\pm 1/2$.
The topological charge density in the intersection plane of three such
configurations is shown in Figs.~\ref{fig:qcontr}a, \ref{fig:qcontr}b and
\ref{fig:qcontr}c. For comparison see the plots in Fig. 2
in~\cite{Hollwieser:2011uj}. If $\phi$ rotates in the $\sigma_1$-subgroup
for the $xy$-vortex and in $\sigma_2$ for the $zt$-vortex then $F\tilde F$
gives no contribution to the topological charge. If we choose the first
vortex sheet of the $zt$-vortex to rotate $\phi$ from zero to $\pi$ in
$\sigma_1$ and on to $2\pi$ in $\sigma_2$ for the second vortex sheet,
whereas the $xy$-vortex only rotates in $\sigma_1$, we get topological
charge $Q_U=1$ with a distribution shown in Fig.~\ref{fig:qcontr}a). If we
rotate $\phi$ for the $xy$-vortex with the exchanged $U(1)$-subgroups as
for the $zt$-vortex from above, {\it i.e.} starting with a $\sigma_2$
rotation and rotating the second vortex sheet in $\sigma_1$ we still get
$Q_U=1$ but now distributed as shown in Fig.~\ref{fig:qcontr}b). Finally,
if we take the $zt$-vortex from above, {\it i.e.} $\sigma_1$ and $\sigma_2$
vortex sheets, and for the $xy$-vortex $\sigma_2$ and $\sigma_3$ vortex
sheets, we find $Q_U=1/2$ distributed as shown in Fig.~\ref{fig:qcontr}c)
with an action density in the intersection plane shown in
Fig.~\ref{fig:qcontr}d).

\begin{figure}
\includegraphics[keepaspectratio,width=\linewidth]{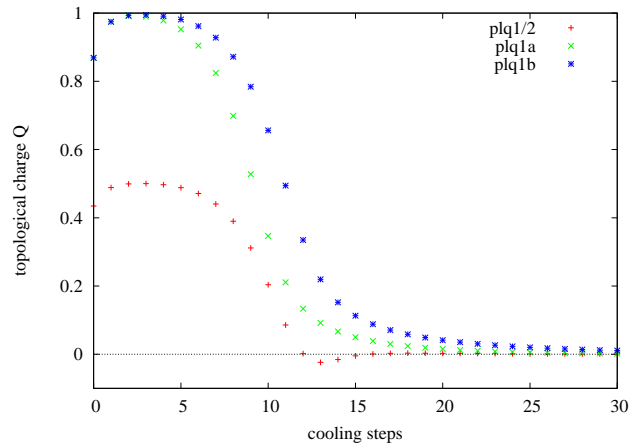}\\
\caption{Topological charge during cooling the (meta-stable) $Q=1$- and $Q=1/2$-configurations of Fig.~\ref{fig:qcontr}. The intersecting vortex sheets end up with anti-parallel orientation and the topological charge vanishes.}
\label{fig:cool1und1/2}
\end{figure}

\begin{figure}
\includegraphics[keepaspectratio,width=\linewidth]{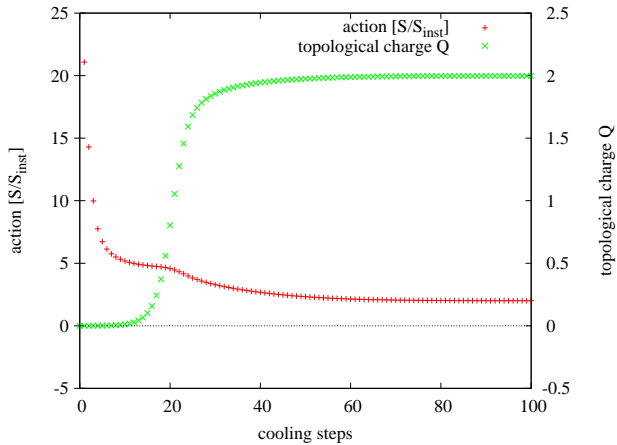}
\caption{During cooling the $Q=0$-configuration the action reaches a plateau with $S=2S_{inst}$ and the parallel vortex pairs end up with parallel color vectors equivalent to the $Q=2$-configuration.}
\label{fig:cool2}
\end{figure}

The action density in Fig.~\ref{fig:qcontr}d) already shows that such orthogonal color vector intersections might be suppressed due to higher action and in fact, at the intersection points we find maximally non-trivial ($yt$-) plaquettes
\begin{align*}
P_{yt}&\approx(-\mathrm{i}\sigma_k)(-\mathrm{i}\sigma_l)\mathrm{i}\sigma_k\mathrm{i}\sigma_l\\
&=(\sigma_k\sigma_l)(\sigma_k\sigma_l)=\mathrm{i}\sigma_m\mathrm{i}\sigma_m=-\mathbbm{1}.
\end{align*}
These rough configurations also seem to trouble the Dirac operators, which do not find any zero modes. During cooling the $Q_U=1$- and $Q_U=1/2$-configurations of Fig.~\ref{fig:qcontr} are only meta-stable and soon turn to anti-parallel vortex pairs with vanishing topological charge, see Fig.~\ref{fig:cool1und1/2}, whereas an original $Q_U=0$-configuration from two parallel vortex pairs in different $U(1)$-subgroups prefers the $S=2S_{inst}$ action minimum and ends up with parallel color vectors yielding $Q_U=2$, see Fig.~\ref{fig:cool2}. For admissible gauge fields (after some cooling) we also find the corresponding numbers of zero modes, {\it i.e.} the different definitions of topological charge agree with each other $Q_U=Q_D$. 

Nevertheless, in the next section we discuss a lattice configuration which immediately guarantees the admissibility condition but gives a discrepancy $Q_U\ne Q_D$.

\section{The "Spherical Vortex Problem"}
\label{sec:sphere}

The spherical vortex of radius $R$ and thickness $\Delta$ was
introduced in~\cite{Jordan:2007ff} and analyzed in more detail
in~\cite{Hollwieser:2010mj}. It is constructed with the following links:
\begin{gather}
U_\mu(x_\nu) = \begin{cases}
 \exp\left(\mathrm i\alpha(|\vec r-\vec r_0|)\vec n\cdot\vec\sigma\right)&t=1,\mu=4\\
 \mathbbm{1}&\mathrm{elsewhere}\end{cases} \label{eq:sphervort}\\ \qquad\mbox{with}\quad \vec n(\vec r,t)=\frac{\vec r-\vec r_0}{|\vec r-\vec r_0|} ~,
\end{gather}
where $\vec r$ is the spatial part of $x_\nu$ and the profile function $\alpha$ is either one of $\alpha_+, \alpha_-$, which are defined by
\begin{gather}
\alpha_+(r) = \begin{cases} 0 & r < R-\frac{\Delta}{2} \\
     \frac{\pi}{2}\left( 1+\frac{r-R}{\frac{\Delta}{2}} \right) &
     R-\frac{\Delta}{2} < r < R+\frac{\Delta}{2} ~~, \\
                        \pi & R+\frac{\Delta}{2} < r
          \end{cases}\\
\alpha_-(r) = \begin{cases} \pi & r < R-\frac{\Delta}{2} \\
     \frac{\pi}{2}\left( 1-\frac{r-R}{\frac{\Delta}{2}} \right) &
      R-\frac{\Delta}{2} < r < R+\frac{\Delta}{2} ~~. \\
                          0 & R+\frac{\Delta}{2} < r
          \end{cases}
\end{gather}
This means that all links are equal to $\mathbbm 1$ except for the
$t$-links in a single time-slice at fixed $t=1$. The phase changes from $0$ to $\pi$ from inside to outside for $\alpha_+(r)$ (or vice versa for $\alpha_-(r)$). The graph of $\alpha_-(r)$ is plotted in Fig. 2 in~\cite{Jordan:2007ff}, giving a hedgehog-like configuration, since  the color vector $\vec n$ points in the ``radial'' direction $\vec r/r$ at the vortex radius $R$. The hedgehog-like structure is crucial for our analysis, leading to the mentioned discrepancy. In maximal center gauge and after center projection, this configuration shows a single, spherical vortex without any intersection or writhing points and hence no topological charge. It is of course possible to construct the same thin spherical vortex after center projection without the hedgehog structure, simply by replacing $\vec n\vec\sigma$ by e.g. $\sigma_3$. Such a vortex has a smooth transition, {\it i.e.,} it is homotopic, to a trivial gauge field. But as in the previous section we are interested in the analysis of topological charge behavior for colorful vortices. 

We would like to underline that for the spherical vortex with the hedge-hog structure there is nowhere a discontinuity in the link variables, not in the center of the three-dimensional spherical vortex nor at the lattice boundary. In both regions the link variables are center elements as necessary for a center vortex. In spite of the hedge-hog structure at the two-dimensional center projected vortex sphere there is no singularity due to the full covering of $S^3$ by the link variables. In other words one can explain the configuration with $t$-links of one time slice rotating from the "south" pole $-\mathbbm 1$ to the "north pole" $+\mathbbm 1$ of $S^3$ in radial direction from the center to the boundary, via color vectors $\vec n\vec\sigma$ given by the spatial components $\vec n=\vec r/r$ of the radius vector. Hence, at the center of the vortex sphere the links belong to the "south pole" of $S^3$, at the two-dimensional vortex surface to the "equator" and at the boundary of the time slice to the "north pole" of $S^3$. In other words, the links $U_t(x,y,z)$ define a smooth non-trivial mapping of $S_3\cong SU(2)$ to $R_3\cup\infty\cong S_3$ which does not contain any singularity. The only singularity that remains in this configuration is the singular gauge transformation leading to the vortex, but such singular gauge transformations are crucial for any vortex structure, as discussed in the introduction.

Now, since only links in the time direction are different from $\mathbbm 1$ for this spherical vortex configuration, the topological charge $Q_U$ determined from any lattice version of $F \tilde F$ vanishes for this spherical vortex configuration. The index of the considered Dirac operators however is nonzero, resulting in $Q_D= \mp 1$, for $\alpha_\pm$. On a $136^3 \times N_t$ lattice the plaquettes for the spherical vortex, Eq.~(\ref{eq:sphervort}), satisfy the ``admissibility'' condition. In fact, the plaquettes get smaller and smaller the bigger the lattice, if we choose $R$ and $\Delta$ to be proportional to the lattice size. We can even get rid of negative links with proper gauges (Landau gauge), ending up in a lattice gauge field with no sign of hiding a singularity at all.

\begin{figure}
\centering
\includegraphics[keepaspectratio,width=\linewidth]{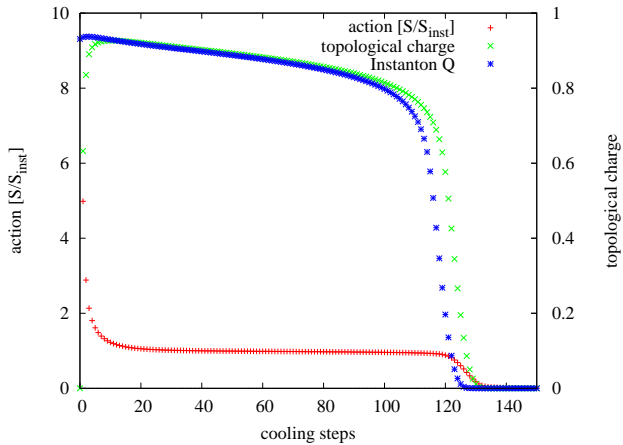}
\caption{Cooling of a spherical vortex on a $40^4$ lattice. The topological charge rises from zero to close to one for $\alpha=\alpha_-$ (right scale) while the action $S$ (in units of the one-instanton action $S_{\rm inst}$) reaches a plateau (left scale). For comparison we also plot the topological charge during cooling of an instanton.}
\label{fig:sphcool}
\end{figure}

In order to understand the discrepancy we apply standard cooling to the spherical vortex configuration. For many cooling steps, the index of the Dirac operator does not change, but the topological charge quickly rises close to $Q_U=\mp 1$ for $\alpha_\pm$  while the action $S$ reaches a (nonzero) plateau. So, the index of the overlap Dirac operator agrees with the topological charge via $F\tilde F$ ($Q_U=Q_D$) after some cooling. In Fig.~\ref{fig:sphcool} we plot the cooling history for a spherical vortex on a $40^4$-lattice. For comparison we also plot the topological charge of an instanton during cooling, which looks pretty much the same as for the spherical vortex. In fact, the action and topological charge densities spread over more and more time-slices, developing a hyper-spherical distribution like standard instantons. We conclude that our spherical vortex develops an instanton-like structure during cooling, in agreement with~\cite{Buividovich:2011mj}, stating that the Hausdorff dimension of regions where the topological charge is localized, tends towards the total space dimensions.

\begin{figure}
\psfrag{x}{$x$}
\psfrag{y}{$y$}
\psfrag{+}{$+Q$}
\psfrag{Q}{}
\begin{tabular}{cc}
a)\includegraphics[keepaspectratio,width=0.45\linewidth]{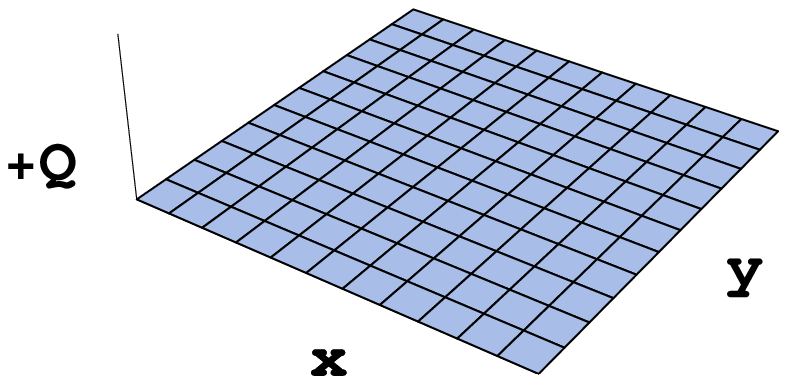} & 
\;b)\includegraphics[keepaspectratio,width=0.45\linewidth]{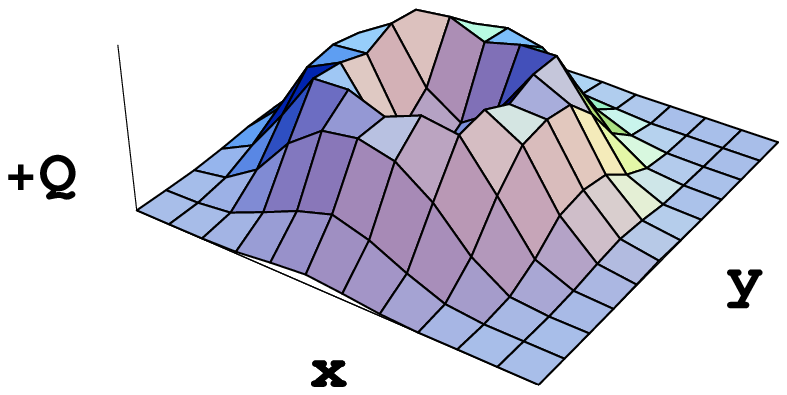}\\ 
c)\includegraphics[keepaspectratio,width=0.45\linewidth]{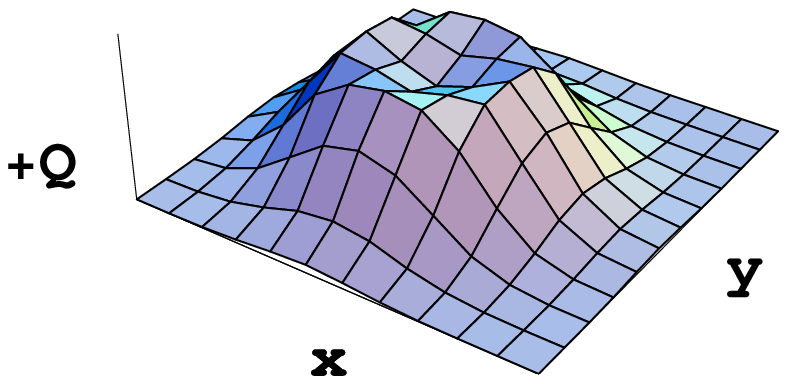} &
\;d)\includegraphics[keepaspectratio,width=0.45\linewidth]{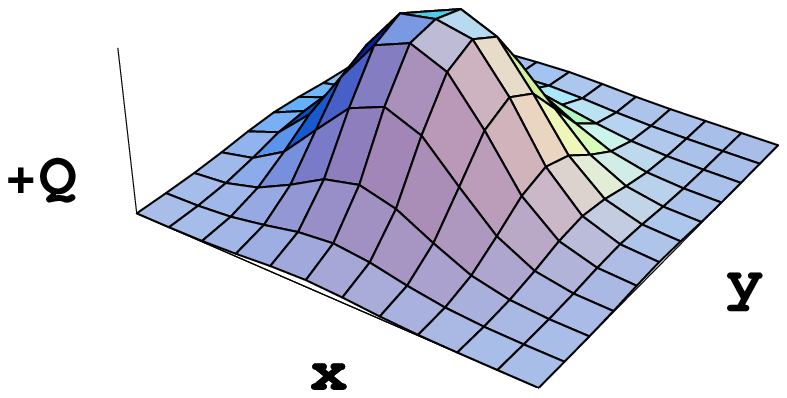} 
\end{tabular}
\caption{Topological charge density via $F\tilde F$ at the $xy$-plane through the center of the spherical vortex during cooling. Before cooling the topological charge $Q_U=0$ (a), evolving
a ring (spherical) distribution along the vortex structure after two cooling steps (b) and developing into an instanton-like (hyper-spherical) distribution after $5$ (c) and $10$ (d) cooling steps.}\label{fig:tdsph}
\end{figure}

However, the vortex structure of our initial configuration is removed after
a few cooling steps, i.e. the spherical vortex shrinks very quickly. This
is in agreement with the fact that the vortex content of a single instanton
is shrunk to a point at the center of the
instanton~\cite{Reinhardt:2001hb,Bruckmann:2003yd}, but it clearly shows
that cooling significantly changes the content of the initial gauge
configuration. In~ (Fig. 5) of \cite{Hollwieser:2010mj} we showed plots of
the shrinking spherical vortex and the corresponding monopole loop in (Fig.
6), as well as the distribution of the scalar density of the zero modes
(Fig. 3), with maxima at the inside (for $\alpha_+$) or the outside (for
$\alpha_-$) of the vortex. In~\cite{Hollwieser:2011uj} we concluded that
Dirac modes are sensitive to Polyakov lines, avoiding negative Polyakov
lines (corresponding to negative links in the construction
in~\cite{Hollwieser:2011uj}). Here, we add some plots of the development of
topological charge density via $F\tilde F$ during cooling in
Fig.~\ref{fig:tdsph}, in order to show that there are no contributions from
any hidden singularities. The topological charge clearly develops from the
vortex surface (ring structure in Fig.~\ref{fig:tdsph}b) shrinking to the
center and developing an instanton-like distribution.

\begin{figure}
\centering
\includegraphics[keepaspectratio,width=\linewidth]{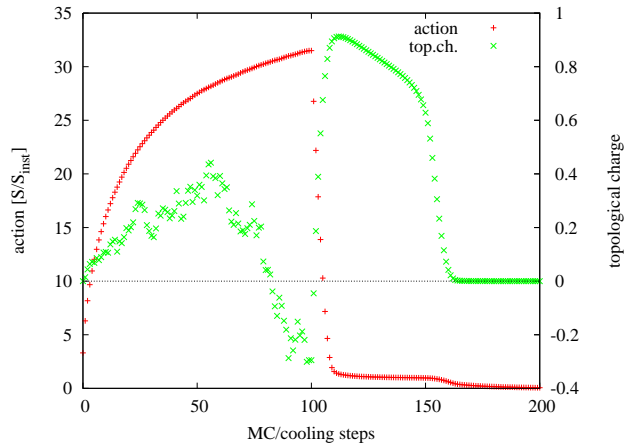}
\caption{Monte Carlo (Metropolis) update and cooling of a spherical vortex on a $16^4$ lattice. After $100$ MC steps, only after some cooling $F\tilde F$ reveals the same topological charge $Q_U=Q_D$ as the Dirac operator.} 
\label{fig:qmc}
\end{figure}

Further we apply some Monte Carlo steps to the spherical vortex configuration using the Metropolis algorithm with a small spread, {\it i.e.} adding small quantum fluctuations, and analyze the vortex structure and topological charge. Fig.~\ref{fig:qmc} shows the action and topological charge during $100$ Metropolis and another $100$ cooling steps. The action rises during the Monte Carlo update and the spherical vortex percolates over the whole lattice while the topological charge fluctuates around zero. The index of the lattice Dirac operator however indicates topological charge $Q_D=\mp1$ for $\alpha_\pm$. Cooling after the Monte Carlo update still reveals the same result for $Q_U=Q_D$ via $F\tilde F$, showing the same behavior as in Fig.~\ref{fig:sphcool}.

\begin{figure}
\centering
\includegraphics[keepaspectratio,width=\linewidth]{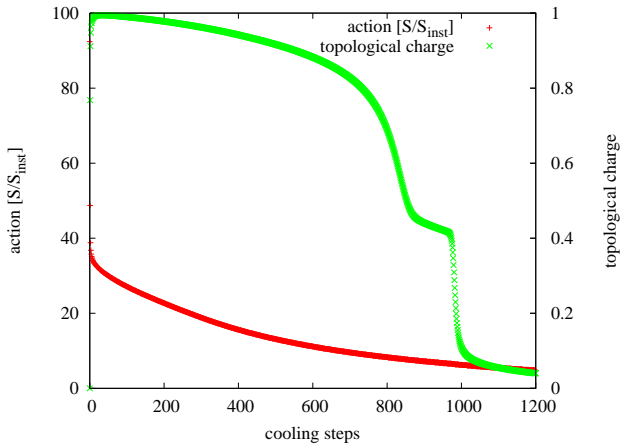}
\caption{Cooling of a spherical vortex on a $136^3\times 2$ lattice. The
topological charge definition $F\tilde F$ initially gives $Q_U=0$, in discrepancy with the index of the Dirac operator. During cooling it rises close to one for $\alpha=\alpha_-$. The second plateau around $900$ cooling steps indicates the fractional topological charge of a Dirac monopole, for more details see~\cite{Hollwieser:2010mj}.}
\label{fig:sphcool136}
\end{figure}

We also want to emphasize the difference to the spherical vortex on an
asymmetric lattice with time extent $N_t=2$, which we analyzed in the second
part of~\cite{Hollwieser:2010mj}. The vortex structure in one of the two
time-slices survives much longer during cooling and leads to a static, singular Dirac monopole before falling through the lattice. For completeness we also plot the cooling history of a spherical vortex on a $136^3\times 2$ lattice in Fig.~\ref{fig:sphcool136}. The configuration initially satisfies the ''admissibility'' condition, but still shows the discrepancy between $F\tilde F$ and the index of the Dirac operator. The second plateau around $900$ cooling steps indicates the fractional topological charge of the Dirac monopole, appearing only on the asymmetric lattice. On a symmetric lattice, we find a continuous transition to an instanton-like structure, and the vortex structure is lost. The singular gauge transformation is smoothed out in time direction at the cost of loosing the center vortex. 


\section{Discussion}\label{sec:disco}
We finally want to resolve the problem of topological charge determination via $F\tilde F$ for the above discussed spherical vortex configuration. The usual expression of topological charge $Q_U$~(\ref{eq:qlatq}) does not take into account the periodic boundary conditions of the lattice, which lead to a different topological classification, even in the large volume limit~\cite{Wrigley:1983pf}. The full expression for $Q$ must contain possible twists in the boundary conditions allowed in the adjoint representation~\cite{'tHooft:1979uj,'tHooft:1981sz,vanBaal:1982ag}. However, such boundary twists may also be hidden in gauge singularities, especially in singular gauge transformations defining center vortices on the lattice. Usually one tries to evaluate the integrals for gauge invariant quantities like~(\ref{eq:qlatq}) in the axial gauge $A_0=0$, where the singularities are transformed away~\cite{Polonyi:1983qv}. However, in our particular case of the spherical vortex there is no way to gauge-transform the singularity away, except by applying the inverse of the initial singular gauge transformation, which then would define the boundary twist, giving the only contribution to the topological charge of our trivial gauge field in $A_0=0$ gauge. 

An easy way of thinking how to determine the topological charge of a gauge field configuration is given in~\cite{Woit:1983tq}. Woit suggests to locate the gauge singularities, locally gauge transform them away using different gauges, and to measure the degrees of the maps relating the different gauges. The sum of these degrees will be a topological invariant, the topological charge. For our configuration we easily may follow the instructions in the above reference by splitting the lattice into time slices. We now can only consider the time slice containing our spherical vortex, where the singular gauge transformation now defines the corresponding mapping and its degree the correct topological charge. 

To state the solution of our problem in a mathematical manner, we consider the homotopy of the above spherical vortex configuration, Eq.~(\ref{eq:sphervort}). The $t$-links of these spherical vortices fix the holonomy of the time-like loops, defining a map $U_t(\vec x,t=1)$ from the $xyz$-hyperplane at $t=1$ to $SU(2)$. Because of the periodic boundary conditions, the time-slice has the topology of a 3-torus. But, actually, we can identify all points in the exterior of the 3 dimensional sphere since the links there are trivial. Thus the topology of the time-slice is $\mathbbm R^3 \cup \{\infty\}$ which is homeomorphic to $S^3$. A map $S^3 \to SU(2)$ is characterized by a winding number 
\begin{gather*}
  N = -\frac{1}{24\pi^2} \int d^3 x
  \,\epsilon_{ijk}\,\mbox{Tr}[(U^\dag\partial_i U)(U^\dag\partial_j
  U)(U^\dag\partial_k U)],\label{eq:winding}
\end{gather*}
resulting in $N=-1$ for positive and $N=+1$ for negative spherical vortices. With this assignment the index of the Dirac operator and the topological charge after cooling coincide with this winding number $N$. Obviously such windings, given by the holonomy of the time-like loops of the spherical vortex, influence the index theorem~\cite{Nye:2000eg,Poppitz:2008hr}, which gives the correct definition of topological charge.

Other, topologically motivated lattice constructions of $Q$ from the gauge field are given in~\cite{Luscher:1981zq} and~\cite{Gockeler:1985dp}, where one compares the gauge rotations necessary in contiguous cells (hypercubes) to put each cell into the same (e.g. axial) gauge. This enables one to construct transition matrices $v_{n,\mu}$ at the lattice sites $n$ common to neighboring cells $c(n)$ and $c(n+\hat\mu)$ which can be used to derive a geometric definition of topological charge 
\begin{align*}
Q & = -\frac{1}{24\pi^2} \sum_{n\in\Lambda} \sum_{\mu,\nu,\rho,\sigma}
\epsilon_{\mu\nu\rho\sigma} \{ \\ & \; 3 \int_{p(n,\mu,\nu)} d^2x \,
\mbox{Tr}[(v_{n,\mu}\partial_\rho
v_{n\mu}^{-1})(v_{n-\hat\mu,\nu}^{-1}\partial_\sigma v_{n-\hat\mu,\nu})] \\ & +
\int_{f(n,\mu)} d^3x \, \mbox{Tr}[(v_{n,\mu}^{-1}\partial_\nu v_{n,\mu})(v_{n,\mu}^{-1}\partial_\rho v_{n,\mu})(v_{n,\mu}^{-1}\partial_\sigma v_{n,\mu})],
\end{align*}
where $\Lambda$ denotes the lattice, $p(n,\mu,\nu)$ the plaquettes and $f(n,\mu)$ the faces (cubes) of a cell $c(n)$. Evaluating this expression for our spherical vortex, the only non-trivial contribution is given by the second term for $f(n,\mu=4)$ in the time-slice of our vortex, resulting in the expression for the winding number given above. All other terms vanish because of trivial transition functions or the vanishing of $F\tilde F$ for our
configuration.

\section{Conclusions}\label{sec:conclusio}
We reported on problems defining topological charge in the background of
classical center vortices on the lattice. First, planar vortex sheets are
constructed by $U(1)$ rotations in a way that they intersect orthogonally. If the gauge
rotations are defined in different $U(1)$-subgroups of $SU(2)$, thus
defining a genuinely non-Abelian configuration, topological charge
contributions at intersection points can occur which are different
from the $Q=\pm1/2$ known to arise in Abelian vortex configurations.
The use of the Dirac operator is not safe in this case because of maximally
non-trivial plaquettes, which seem to suppress such configurations in the
functional integral. However, for ``admissible'' gauge configurations of
classical, spherical center vortices we find a discrepancy between $F\tilde F$
and the lattice index theorem, for both, overlap and asqtad staggered fermions
in the fundamental and adjoint representations. Numerically, the discrepancy
equals the winding number of the spheres when they are regarded as maps $S^3 \to
SU(2)$. The problem arises due to the periodic boundary conditions on the
lattice and the fact that the center vortex configuration under consideration is
based on a singular gauge transformation. In our case we can regard the singular
gauge transformation resulting in our spherical vortex as a boundary twist,
leading to an extra contribution to $F\tilde F$, resolving the discrepancy.
However, this result shows that the interpretation of topological charge via
$F\tilde F$ is rather subtle in the background of center vortices. The mentioned
admissibility conditions do not guarantee that a naive plaquette or hypercube
definition of topological charge gives a result which accords with a counting of
Dirac zero modes. This is only guaranteed in the continuum limit, where vortex configurations may become singular and validate the derivation of topological charge via $F\tilde F$. On lattice configurations the $F\tilde F$ definition of topological charge should only be used after cooling, which however may change the gauge field content significantly.

\acknowledgments{We are grateful to Martin L\"uscher, {\v S}tefan Olejn\'{\i}k and Mithat \"Unsal for helpful discussions. This research was partially supported by the Austrian Science Fund (``Fonds zur F\"orderung der Wissenschaften'', FWF) under contract P22270-N16 (R.H.).}

\bibliographystyle{unsrt}
\bibliography{../literatur}

\end{document}